\documentclass[aps,prd,preprint,groupedaddress,showkeys]{revtex4-1}

\usepackage{graphicx,overpic,color}

\begin{document}


\title{Probing the internal structure of baryons} 


\author{Guangshun Huang$^{1,2}$}
\email[]{hgs@ustc.edu.cn}
\affiliation{
$^{1}$Department of Modern Physics, 
University of Science and Technology of China, 
Hefei 230026, People’s Republic of China\\
$^{2}$State Key Laboratory of Particle Detection and Electronics, 
Hefei 230026, People’s Republic of China}

\author{Rinaldo Baldini Ferroli$^{3,4}$}
\email[]{baldini@lnf.infn.it}
\affiliation{
$^{3}$INFN Laboratori Nazionali di Frascati, I-00044 Frascati, Italy\\
$^{4}$Institute of High Energy Physics, 
Beijing 100049, People’s Republic of China}

\collaboration{BESIII Collaboration}


\begin{abstract}
Electromagnetic form factors are fundamental observables
that describe the electric and magnetic structure of hadrons 
and provide keys to understand the strong interaction.
At the Beijing Spectrometer (BESIII), form factors have been 
measured for different baryons in the time-like region
for the first time or with the best precision.
The results are presented with examples focus on but not limited to 
the proton/neutron, the $\Lambda$, with a strange quark, 
and the $\Lambda_c$, with a charm quark.
\end{abstract}


\maketitle

\section{Introduction}

Baryons and mesons are both hadrons, {\it i.e.}, 
bound systems of quarks in naive quark model~\cite{quarkmodel} 
or more accurately also gluons in modern theory.
Baryons are half-integer spin fermions, comprised, in a first approximation, 
of 3 quarks held together by the strong interactions. 
Protons ($p$) and neutrons ($n$), collectively known as nucleons ($N$), 
are the lightest baryons, and are the major components of 
the observable matter of the Universe.
A nucleon has three valence light quarks ($u$ or $d$); 
if one or more of its $u$ or $d$ quarks are replaced by 
heavier quarks ($s$, $c$, $b$ or $t$), it becomes a hyperon. 
The most known baryons are the spin 1/2 SU(3) octet, including 
isospin doublet $p/n$, singly-stranged isospin singlet $\Lambda$,
singly-stranged isospin triplet $\Sigma^-/\Sigma^0/\Sigma^+$
and doubly-stranged isospin doublet $\Xi^-/\Xi^0$~\cite{PDG2020}.
The lightest charmed baryon is the $\Lambda_c^+$~\cite{PDG2020}.
Hadrons are not point-like particles, and their internal electric 
and magnetic structure is characterized by their electromagnetic 
form factors (FF).

\begin{figure}[tb]
\begin{center}
\includegraphics[width=0.8\textwidth]{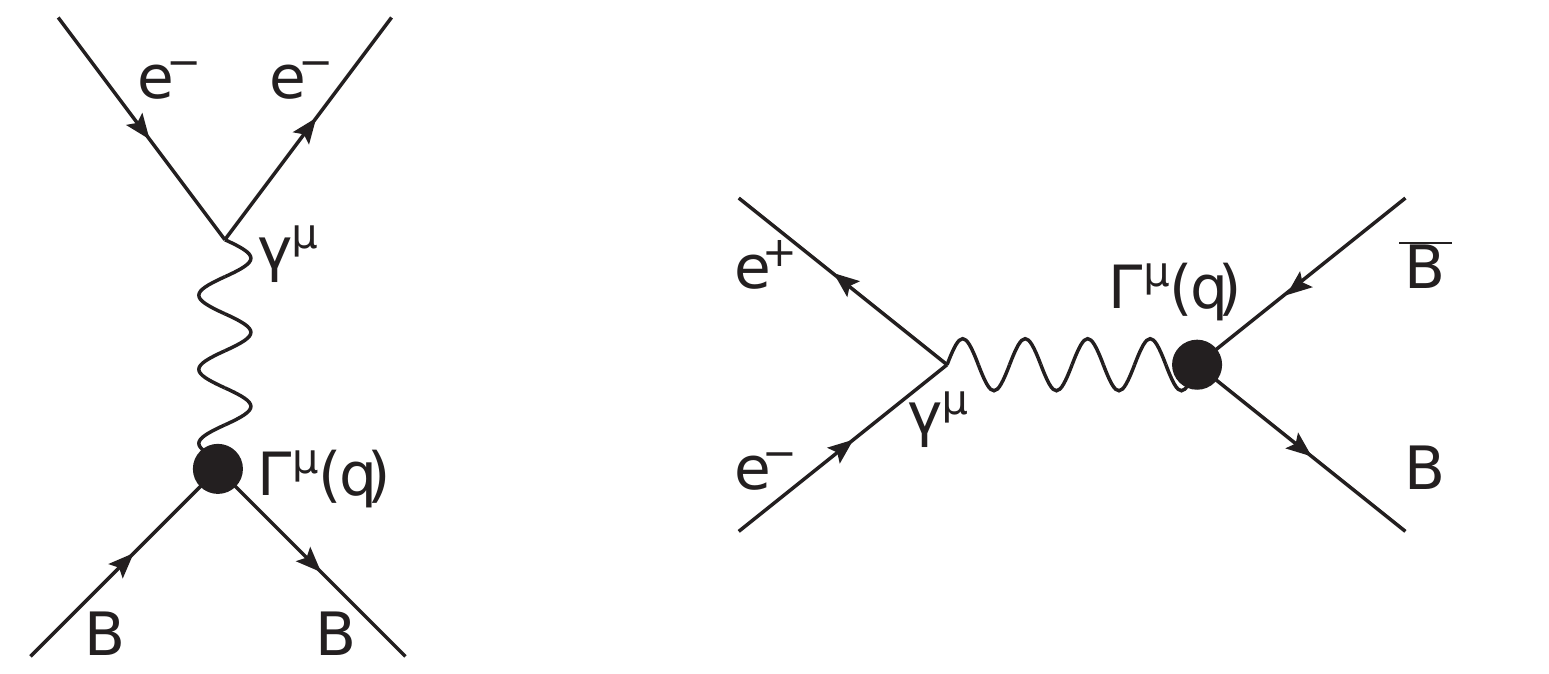}
\put(-280,130){(a)}\put(-50,130){(b)}
\end{center}
\caption{Lowest-order Feynman diagrams for elastic electron-hadron 
scattering \mbox{$e^- B  \rightarrow e^- B$} (a), and for the 
annihilation process \mbox{$e^+e^- \rightarrow B \bar{B}$} (b).}
\label{Feynman}
\end{figure}

The particles are so tiny (at the order of 10$^{-15}$ m, or fm) that they 
cannot be observed directly by the human eye (ability of 10$^{-4}$ m, or 0.1 mm), 
an optical microscope (resolution of 10$^{-7}$ m, or 0.1 $\mu$m), or even 
an electric microscope (resolution of 10$^{-10}$ m, or 0.1 nm, size of an atom). 
Instead, their properties are studied through collisions. 
When two particles traverse each other, they interact by 
exchanging force carriers called bosons that
transfer some energy and momentum ({\it i.e.}, four momentum) 
from one to the other.
For electron-nucleon scattering, the electron is a probe that spies on
the secrets hidden inside the nucleon, and in this case the four-momentum 
transfer squared has a negative value ($q^2<0$), 
and is categorized as a space-like process.
When a particle and an anti-particle meet, for example the case
of an electron and a positron, they can annihilate -
{\it i.e.}, disappear into a virtual photon, and then produce
fermion-antifermion pair that eventually materializes as 
a system of hadrons, of which a baryon-antibaryon pair is one possibility. 
In this case the four-momentum transfer squared 
has a positive value ($q^2>0$), and is classified as a time-like process. 
The Feynman diagrams for these two processes are shown in 
Fig.~\ref{Feynman} (a) and (b), respectively.
For the latter, the form factors of the participating baryon can be deduced 
from the behavior of the outgoing baryon-antibaryon pair, 
which is subject of the study covered in this paper.

Hadronic production data from electron-positron annihilations
at low energies (around GeV order) is important input to 
the understanding of the structure of hadrons and 
the strong interactions of their constituent quarks.
Moreover, since hyperons are not stable, they can be studied 
only in the time-like domain.
The Beijing Spectrometer (BESIII)~\cite{BES3} at the Beijng Electron
Positron Collider (BEPCII)~\cite{BEPC2} operates in the center-of-mass 
energy range from 2.0 and 4.6 GeV, which is a transition region 
between perturbative and non-perturbative Quantum Chromodynamics (QCD).
Using the initial state radiation (ISR) technique, 
BESIII can also access energies below 2.0 GeV.
The $e^+e^-$ collision data that is used for QCD studies at BESIII include 
an integrated luminosity of 12 pb$^{-1}$ at four energies 
(2.23, 2.4, 2.8 and 3.4 GeV) in the continuum taken in 2012, 
about 800 pb$^{-1}$ at 104 energies between 3.85 and 4.6 GeV taken 
in the 2013-2014 run, and about 650 pb$^{-1}$ at 22 energies 
from 2.0 to 3.08 GeV taken in 2015.
These are the so-called scan data, with moderate luminosity 
at each energy point, nonetheless, for these energies 
they are the largest data samples in the world.
There are also much larger samples for charm physics or XYZ particle 
search, some as large as a few fb$^{-1}$'s at a single energy,
which are suitable for ISR-type analyses.
With these huge data samples, BESIII is uniquely well suited to make 
baryon form factor measurements with unprecedentedly high precision.

\vspace{-0.8cm}

\section{Baryon Mysteries}
The standard wisdom is that baryons are bound states of three quarks,
but this description is incomplete. 
For example, though nucleons are the basic building blocks of 
observable matter in the Universe, not all of their basic properties 
such as their size, spin, magnetic moment and mass are fully 
understood, even after 100 years of study~\cite{Rutherford, Chadwick}.

The charge radius of proton measured by muonic Lamb shift
once differed from that determined by electron-proton scattering or 
electronic Lamb shift as large as five standard deviations~\cite{Pohl},
but recent measurements from electron scattering~\cite{PRad} 
and hydrogen spectroscopy~\cite{elLamb} 
eliminated the discrepancies, and this so-called proton-radius puzzle 
has been essentially solved~\cite{SciBull, NatureRevPhys}.

The proton spin has also been in a crisis in the era of the 
constituent quark model. 
The European Muon Collaboration (EMC) experiment found that
the baryon spin is not only due to the spins of 
the valence quark~\cite{SpinAsym}.
It has been commonly assumed that the proton's spin=1/2 
was formed by two quarks with parallel spins and
the third quark with opposite spin.
In the EMC experiment, a quark of a polarized proton target was struck
by a polarized muon beam, and the quark's instantaneous spin was measured. 
It was expected that the spin of two of the three quarks would cancel out 
and the spin of the third quark would be polarized in the direction of the
proton's spin. Thus, the sum of the quarks' spin was expected to be
equal to the proton's spin.
Surprisingly, it was found that the number of quarks with spin in the 
proton's spin direction was almost the same as the number of quarks 
whose spin was in the opposite direction.
Similar results have been obtained in many experiments afterwards,
demonstrating clearly that both generalized parton distributions 
and transverse momentum distributions are important 
in the nucleon spin structure~\cite{NucleonSpin}.
Our modern understanding is that the nucleon spin comes 
not only from quarks but also from gluons, 
and various contributions can be calculated using 
e.g. Ji's sum rule~\cite{Ji}.
The abnormal magnetic moment of proton (much larger than
that for a Dirac point-like particle) is generally considered as
an indication of a more complicated internal structure
than simply three spin=1/2 quarks in a relative $S$-wave.

Moreover, the mass of proton also cannot be explained by Higgs mechanism, 
since the sum of mass of quarks inside a proton is too small,
which means there are considerable contributions to its mass from
the strong interactions among quarks and gluons.
Nowadays these contributions can be calculated precisely in the lattice QCD, 
so the proton mass is largely understood~\cite{sigmaterm, npmass}.

\section{Baryon form factor measurements at BESIII}
The differential cross section of electron-positron annihilation
to a baryon-antibaryon pair can be written as a function 
of the center-of-mass (c.m.) energy squared $s$ as~\cite{Zichichi},
\begin{equation}
\label{eq:xsBB}
\frac{d\sigma_{B\bar{B}} (s)}{d\Omega}
=\frac{\alpha^{2}\beta C}{4s}\Bigl[|G_{M}(s)|^2(1+\cos^{2}\theta)
+\frac{4m_{B}^{2}}{s}|G_{E}(s)|^2\sin^{2}\theta\Bigr],
\end{equation}
where $\theta$ is the polar angle of the baryon in the $e^+e^-$ c.m.~frame 
and $\beta = \sqrt{1-4m_{B}^2/s}$ is the speed of baryon.
The Gamov-Sommerfeld factor $C$~\cite{Sommerfeld, Brodsky, Sakharov} 
describes the Coulomb enhancement effect: for charged baryon pair 
$C = y/(1-\mathrm{e}^{-y})$ with $y = \pi \alpha\sqrt{1-\beta^{2}} / \beta$, 
accounts for the electromagnetic interaction between the outgoing baryons;
while for neutral baryon pair $C=1$.
The form factors, $G_E$ and $G_M$, essentially describe the 
electric and magnetic distributions inside the baryon and 
basically provide a measure of its boundary or size.
These are functions of the four-momentum transfer $s=q^2$,
so more accurately should be written as $G_E(q^2)$ and $G_M(q^2)$.
In the time-like domain, the form factors are complex
with a nonzero imaginary part, and the translation into the 
internal structure is not straightforward, in contrary to 
the case in the space-like region. It is noteworthy that 
final-state interactions become prevailing close to threshold 
and thus should be properly dealt with.
By definition the electric and the magnetic form factors 
are equal at the baryon-antibaryon pair's mass threshold
where only $s$-wave production contributes~\cite{Griffiths}, 
{\it i.e.}, $G_E(4m_B^2)=G_M(4m_B^2)$,
but generally they are not.
In analyses of data with limited statistics it is often 
assumed that they are equal and the two form factors are replaced 
by an effective form factor, $G_{eff}=G_E=G_M$.

In principle, the Coulomb interaction between the outgoing 
charged baryon-pair $B^+B^-$ should play an important role, 
in particular by producing an abrupt jump in the cross section 
at threshold, since the phase space factor $\beta$ is cancelled by a 
$1/\beta$ factor in the Coulomb correction (however there is no 
full consensus on that), which is a non-perturbative correction to 
the Born approximation to account for the Coulomb interaction between 
the outgoing charged baryons. In fact, the cross section for the 
$e^+e^- \to p\bar{p}$ at threshold has been measured to be very close 
to the pointlike value, which is consistent with the prediction,
but then it is followed by a flat behavior, which is unexpected. 
While for a neutral-baryon-pair $B^0\bar{B}^0$, the cross section
at threshold should be zero according to Eqn.~\ref{eq:xsBB}.
The minimum c.m. energy for BESIII data is 2.0 GeV, which is 
about 122 MeV above the nucleon-antinucleon threshold, so no solid 
conclusion can be drawn for the proton-pair and neutron-pair cases, 
but BESIII can test these effects for charged baryons by seeing 
if there is a step with a value close to the pointlike one 
for $\Lambda_c^+\bar{\Lambda}_c^-$ production, 
and for neutral baryons by seeing if the cross section is vanishing 
at the $\Lambda\bar{\Lambda}$ at threshold. 
Present BESIII results seem to indicate that at both the 
$\Lambda_c^+\bar{\Lambda}_c^-$ and $\Lambda\bar{\Lambda}$ thresholds 
there is a step that is close to the pointlike value for charged particles, 
although maybe not exactly the same. 

\subsection{Proton}
\begin{figure}[tp]
 \centering
 \includegraphics[height=5.5cm]{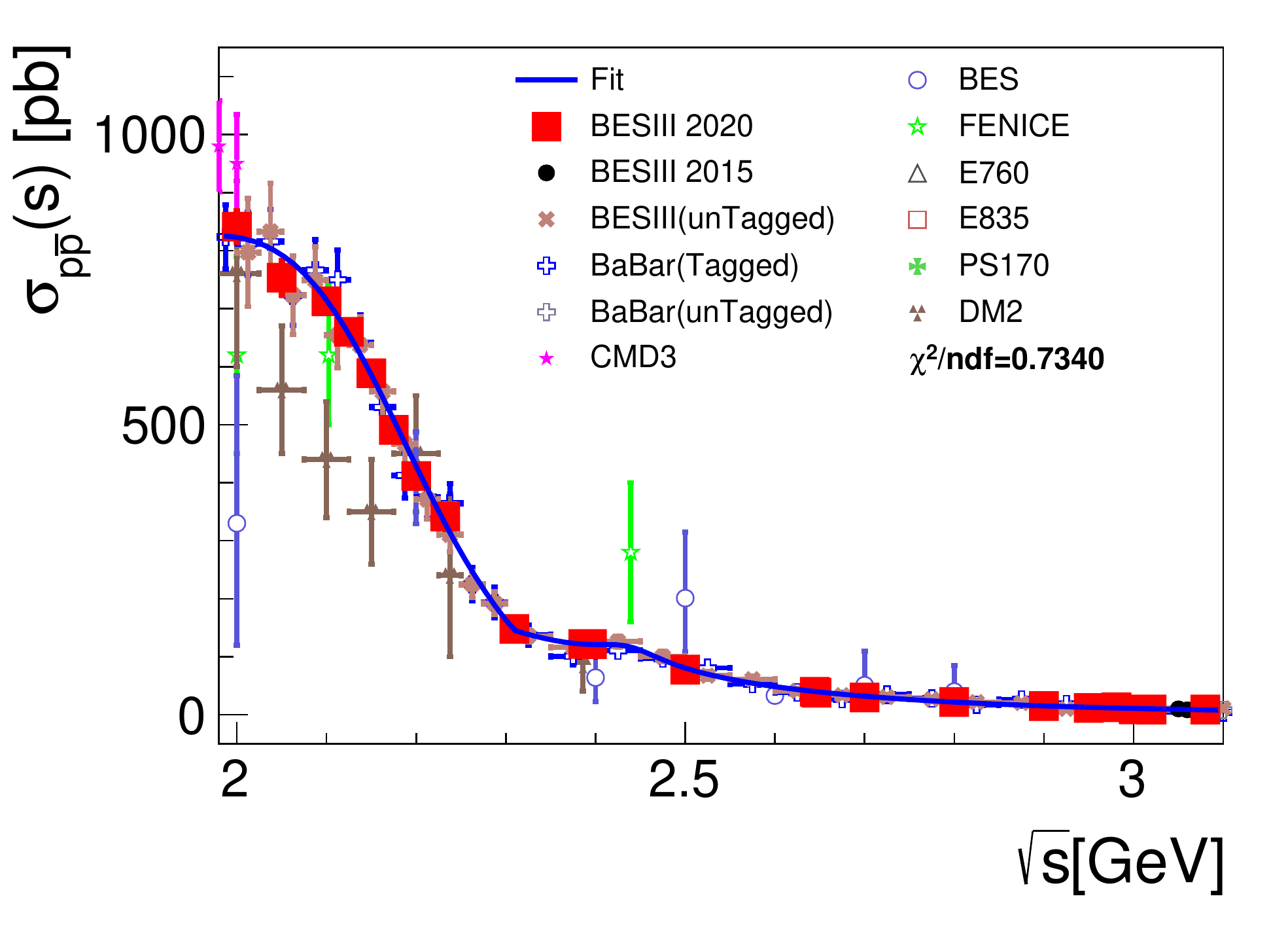}\hspace{0.5cm}%
 \includegraphics[height=5.5cm]{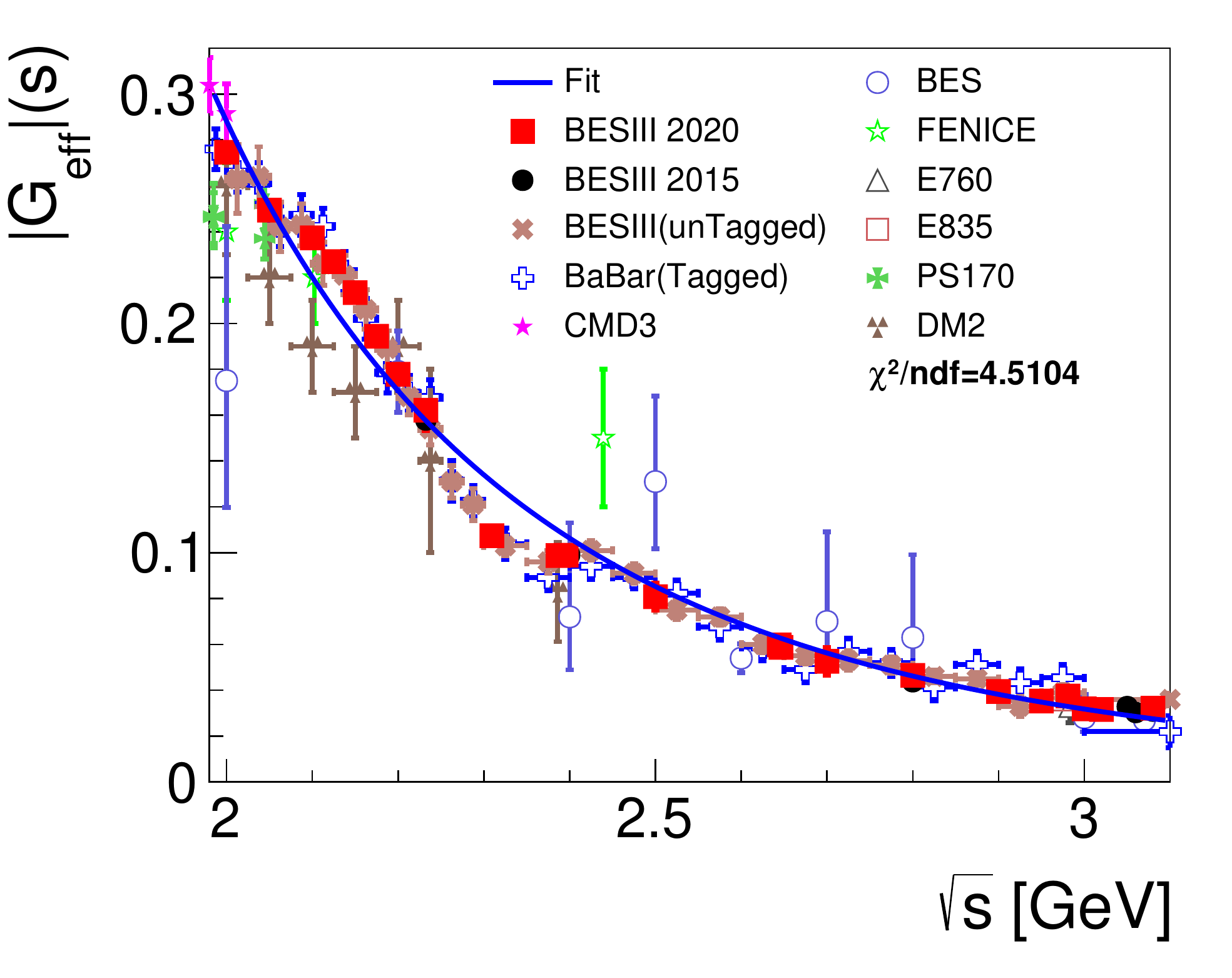}\hspace{0.5cm}%
 \put(-375,135){(a)}\put(-150,135){(b)}\\
 \includegraphics[height=5.5cm]{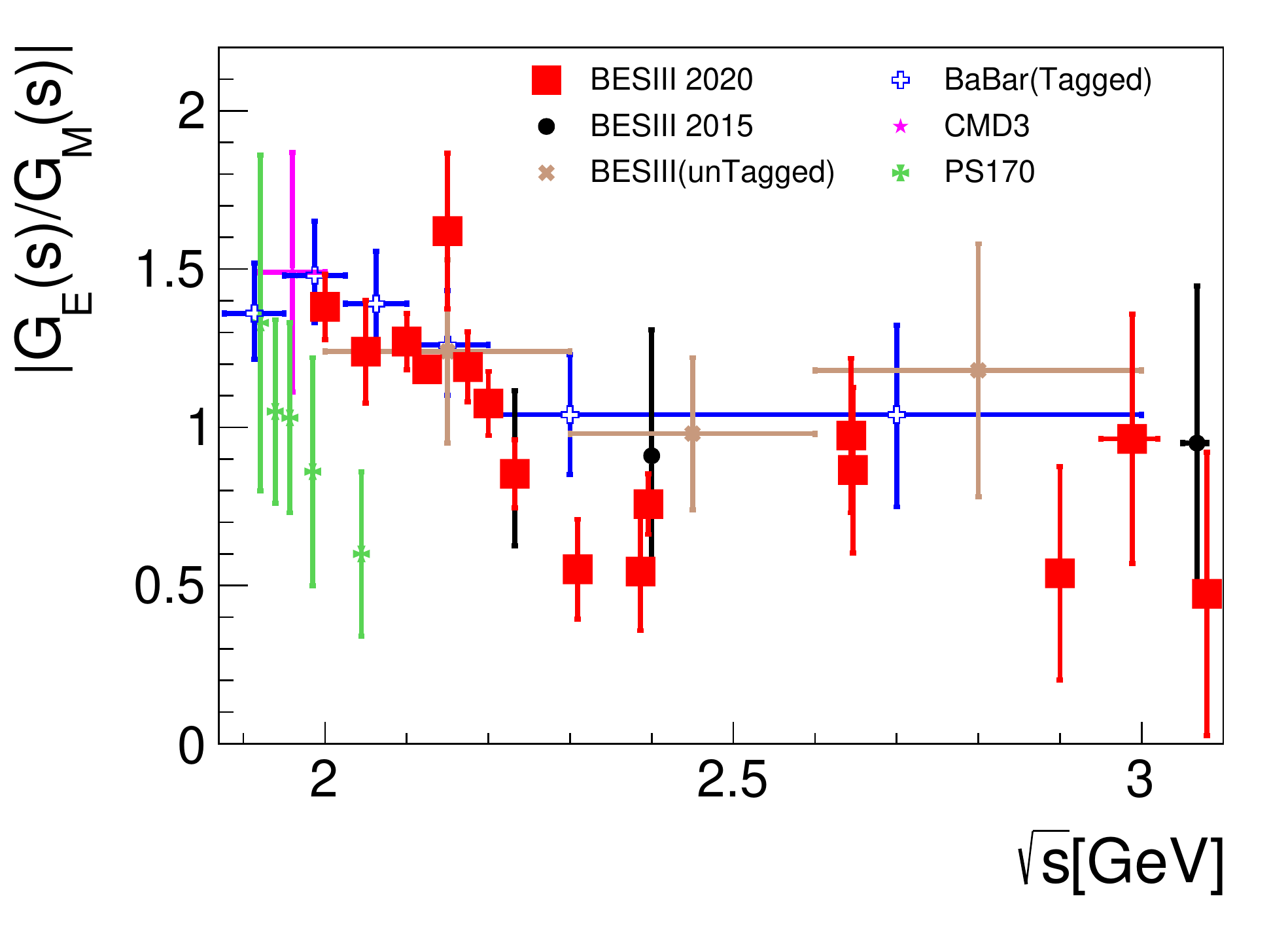}\hspace{0.5cm}%
 \includegraphics[height=5.5cm]{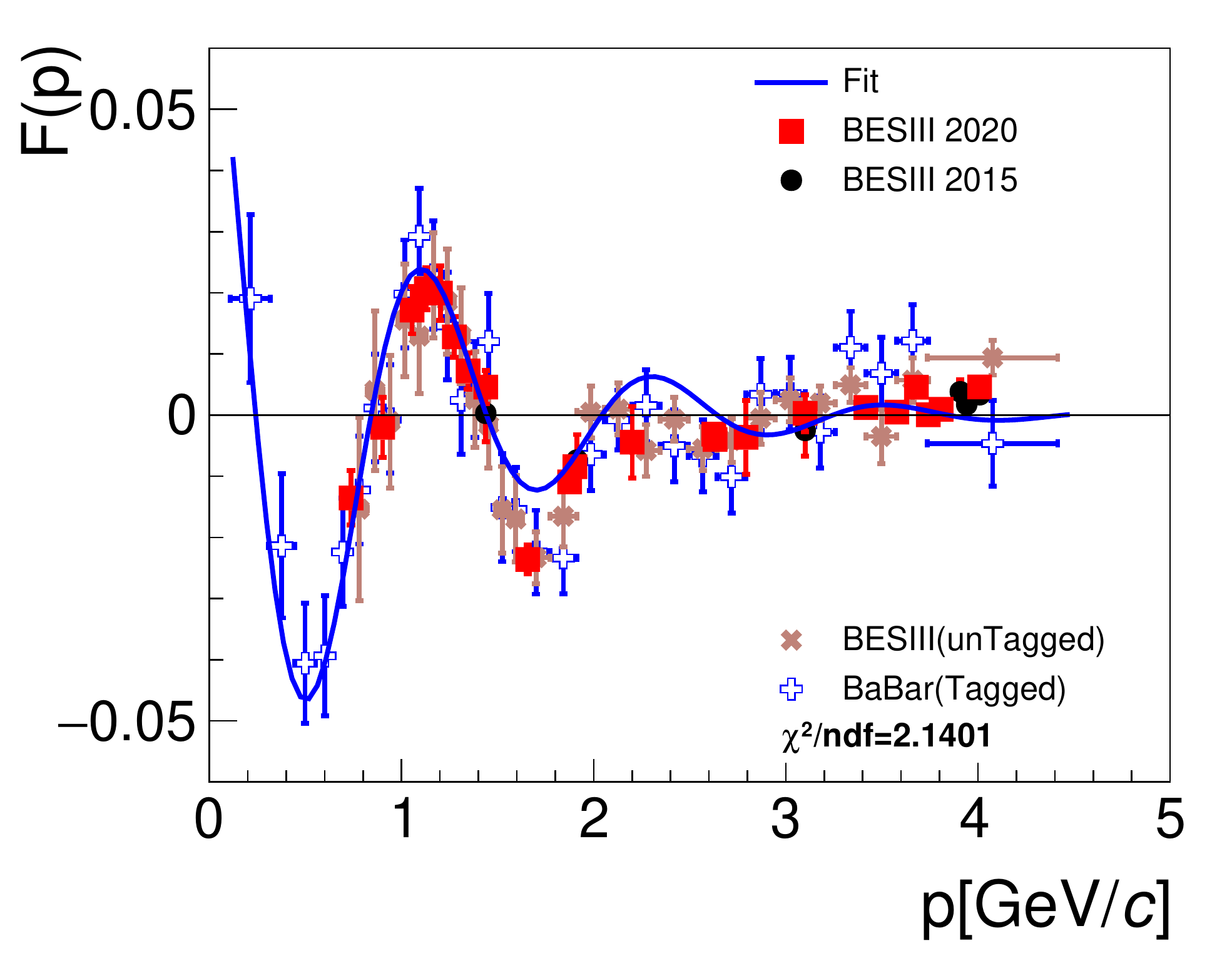}
 \put(-375,135){(c)}\put(-150,135){(d)}
 \caption{\label{fig:proton}
(a) the cross sections for $e^+e^- \to p\bar{p}$ cross section. 
(b) the effective proton time-like form factor. 
The blue curve is the results of an attempt to fit the measurements 
with smooth dipole-like function.
(c) the ratio $R=|G_E/G_M|$. 
(d) effective form factor residual $F(p)$ after subtracting 
the one calculated by QCD theory (the blue curve shown in (b)), 
as a function of the relative motion $p$ of the final proton and antiproton.
Plots are from Ref.~\cite{ppbar2020}.}
\end{figure}

Space-like proton form factors have been measured 
with very high precision by many experiments~\cite{JLab1, JLab2}.
In the time-like region, there have been a few measurements 
of $G_{eff}$, by DM2~\cite{DM2pp, DM2BB}, E760~\cite{E760}, 
PS170~\cite{PS170}, FENICE~\cite{FENICE}, E835~\cite{E835a, E835b}, 
BaBar~\cite{ppbarBaBar1, ppbarBaBar2} and CMD-3~\cite{ppCMD3, CMD3}, 
but these have relatively poor precision and mutual agreement.
For $|G_E/G_M|$ ratio, the measurements were rare and 
there is a long-time tension between PS170 and BaBar.
The BESII experiment also measured the proton effective form factor,
but with poor statistical precision~\cite{lihh}.
BESIII continued this effort using the 2012 and 2015 scan data, 
and produced the most accurate $|G_E/G_M|$ ratio measurements 
at 16 c.m. energies between 2.0 and 3.08 GeV~\cite{ppbar2015, ppbar2020} 
that favor BaBar over PS170 and helped clarifying the puzzle. 
BESIII also performed the measurements using the ISR 
technique~\cite{ppbaruntag, ppbartag}, 
with results that are consistent with BaBar's.
The BESIII measurements are shown in Fig.~\ref{fig:proton} 
(a) for $p\bar{p}$ production cross section in $2.0 - 3.08$ GeV,
(b) the effective proton time-like form factor,
(c) the form factor ratio $R=|G_E/G_M|$, and
(d) the effective form factor residual,
together with results from other experiments.
The best precision in the time-like region was reported by BESIII, 
and the electric form factor was extracted for the first time.
The unprecedented 3.5\% uncertainty that was achieved at 2.125 GeV 
by BESIII is close to that of the best measurements in the space-like 
region, which have been at per cent level for a long time.
The CMD-3 experiment measured the production cross section of proton pair 
and observed an abrupt rise at the nucleon-antinucleon threshold~\cite{CMD3},
as expected for point-like charged particles according to Eqn.~\ref{eq:xsBB}. 
BESIII did not extend down to the threshold energy, but the results 
around 2 GeV agree with CMD-3.
This information improves our understanding of the proton inner structure 
from a different dimension and helps to test theoretical models 
that depend on non-perturbative QCD, e.g. charge distribution 
within the proton can be deduced~\cite{unitary1, unitary2}.
The near threshold behavior of the electromagnetic form factor 
of a hadron is mostly determined by the interaction of the 
hadron-antihadron in the final state, and therefore the measurements 
of the form factor properties can also serve as a fruitful source of 
information about hadron-antihadron interaction~\cite{Dalkarov}.

Interestingly there are oscillations in the effective proton form factor, 
first seen by BaBar and later confirmed by BESIII~\cite{ppbaruntag}.
These oscillations were subsequently studied with more precise data
by BESIII~\cite{ppbar2020}.
Ref.~\cite{PeriodicFF} speculated that possible origins of this 
curious behavior are rescattering processes at relative distances 
of 0.7 - 1.5 fm between the centers of the forming hadrons, leading 
to a large fraction of inelastic processes in $p-\bar{p}$ interactions, 
and a large imaginary component to the rescattering processes.

\subsection{Neutron} 

\begin{figure}[htbp]
\begin{center}
\includegraphics[width=0.6\textwidth]{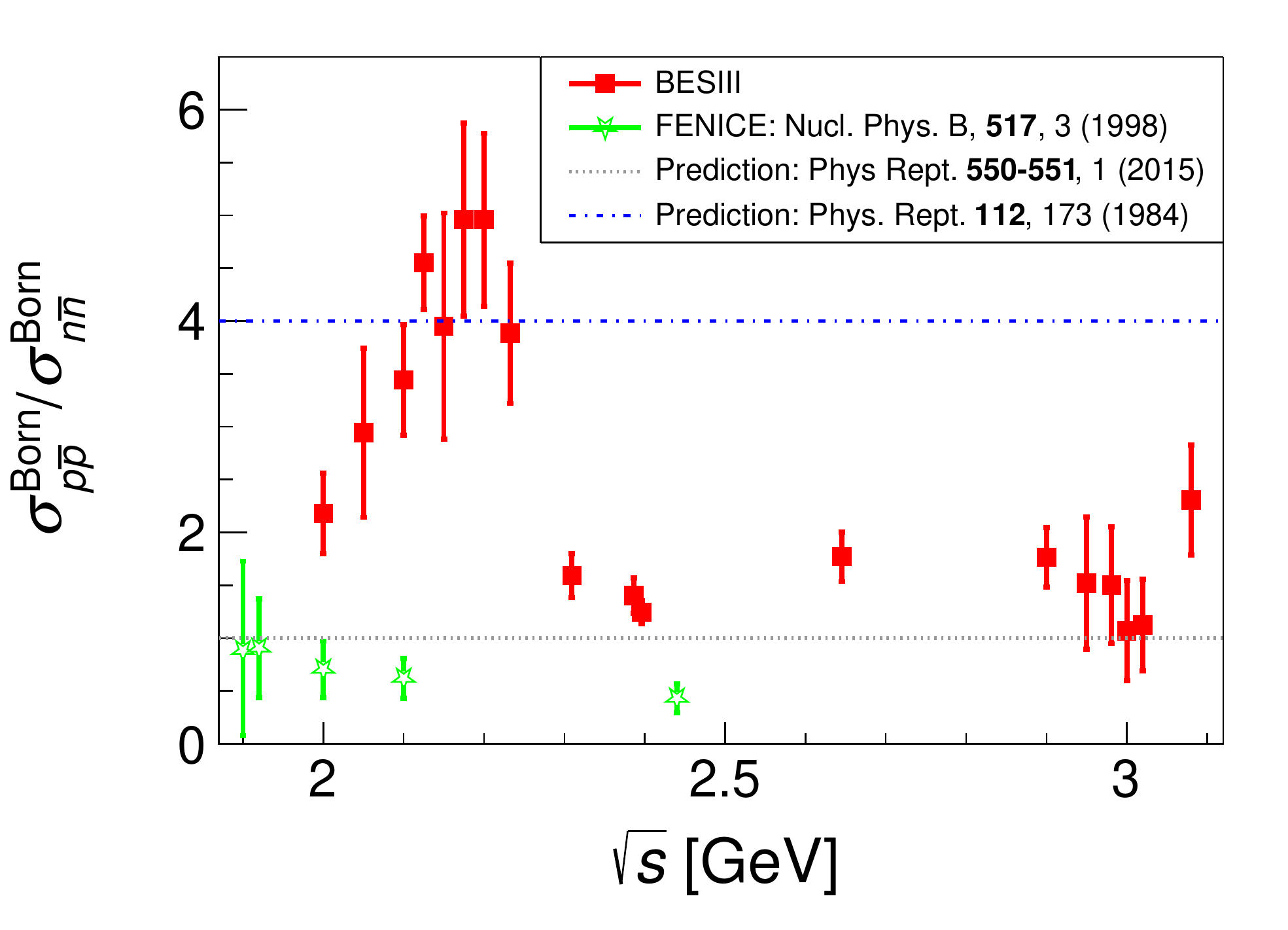}
\caption{Ratio of Born cross section of $e^+e^- \to p\bar{p}$
to that of $e^+e^- \to n\bar{n}$.}
\label{pp2nn}
\end{center}
\end{figure}

Prior to the BESIII experiment, there was a long standing puzzle 
related to differences between the neutron and proton production rates. 
QCD-motivated models predict that the cross section for the proton
should be 4 times larger than for neutron~\cite{QCDcalc},
or they should be same~\cite{Pacetti}. 
In contrast the FENICE experiment found that the neutron cross section 
was twice as large as the the proton's, albeit with statistics 
that were very limited, only 74 $n\bar{n}$ events in total 
for five energy bins~\cite{FENICE}.
More recent measurements in the vicinity of the nucleon-antinucleon 
threshold are from the SND experiment~\cite{SND2014, SNDnn}.
The cross sections of $e^+e^- \to n\bar{n}$ and 
the neutron form factors between 2 GeV and 3.08 GeV
have been measured by BESIII with a good deal more data, 
over 2000 $n\bar{n}$ events at 18 energies~\cite{nnbar}. 
Because the final state neutron and anti-neutron
are both neutral, with no tracks recorded in
the drift chamber, the event selection is a challenge.
The information in the calorimeter and the 
time of flight counters has to be used to identify the signal;
as such the selection efficiency is much lower and  
the number of observed neutron events is significantly 
less than that for protons. 
Neutron measurements from SND~\cite{SND2014, SNDnn} and BESIII~\cite{nnbar}
overlap and roughly agree at 2 GeV, where a cross-section behavior 
that is close to the $e^+e^- \to p\bar{p}$ case is observed,
in particular a flat behavior above threshold up to 2 GeV 
as seen by CMD-3~\cite{CMD3}, but this challenges 
the expected behavior from Eqn.~\ref{eq:xsBB}.
For energies above 2 GeV, the BESIII measurements of the
ratio of the proton to neutron cross sections is 
more compatible with the QCD-motivated model predictions:
as shown in Fig.~\ref{pp2nn}, the cross section for $e^+e^- \to p\bar{p}$ 
is larger than for $e^+e^- \to n\bar{n}$ in general.

From BESIII measurements of the angular distributions for 
$e^+e^- \to N\bar{N}$ events, the $S$-wave and $D$-wave contributions 
are disentangled for the first time, 
that is currently under further investigation in the Collaboration.
Moreover from comparisons of the $e^+e^- \to n\bar{n}$ and 
$e^+e^- \to p\bar{p}$ cross sections, the isoscalar and isovector 
components of $e^+e^- \to N\bar{N}$ can, in principle, 
be separated~\cite{IsoDomi}. 
One of the components dominates and is nearly constant up to 2 GeV, 
similar to $e^+e^- \to p\bar{p}$, but at present it is difficult 
to identify whether the isoscalar (very likely the largest) or 
the isovector one. With more data in the future, this identification 
could be achieved by BESIII.

\subsection{The $\Lambda$ hyperon}
\begin{figure}[t!]
\begin{center}
\begin{minipage}{0.45\linewidth}
\centerline{\includegraphics[width=1\linewidth]{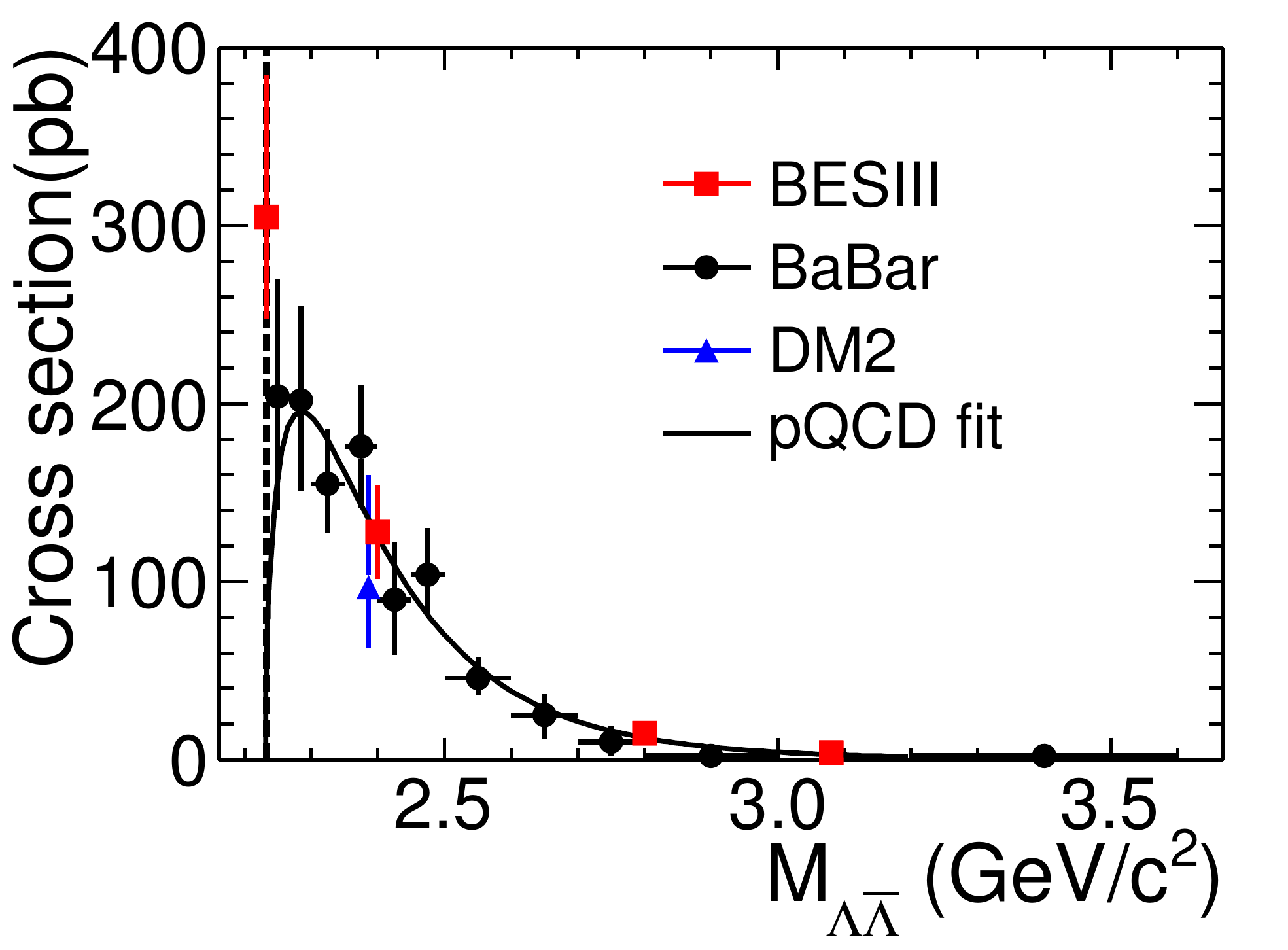}}
\end{minipage}
\put(-150,55){(a)}
\begin{minipage}{0.45\linewidth}
\centerline{\includegraphics[width=1\linewidth]{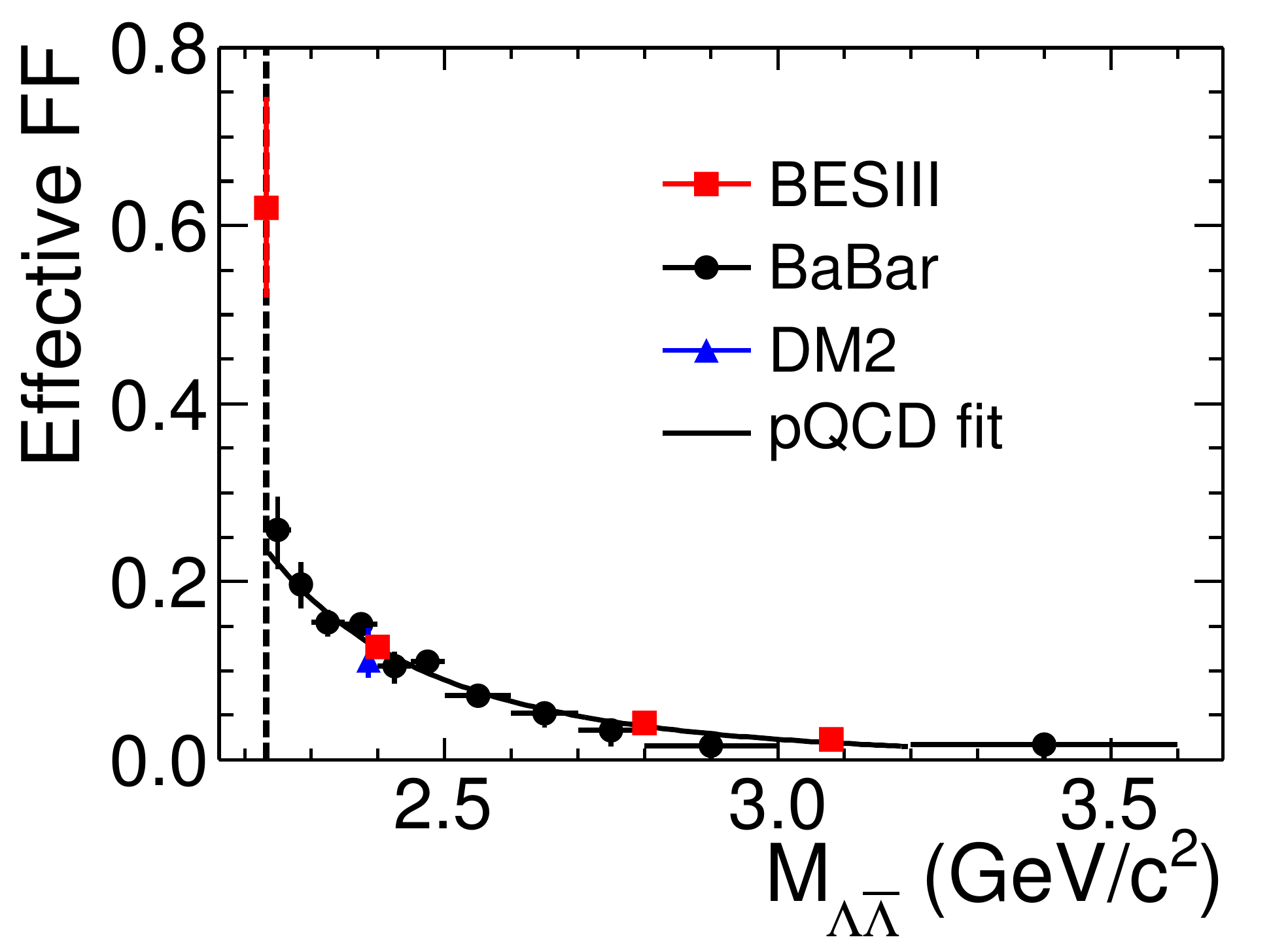}}
\end{minipage}
\put(-150,55){(b)}
\end{center}
\caption[]{(a) Measurements of $e^+e^{-} \rightarrow \Lambda \bar{\Lambda}$
cross section. (b) $\Lambda$ effective form factor.
Plots are from Ref.~\cite{Lambdapair}.}
\label{lambda}
\end{figure}

The $\Lambda$, which is the lightest hyperon that contains an $s$ quark, 
is more difficult to study than the nucleon because of its smaller 
production cross section.
It was measured previously by the DM2~\cite{DM2BB} and
BaBar~\cite{LambdaBaBar} experiments, but with results were not conclusive.
BESIII has studied the channel $e^+e^{-} \rightarrow
\Lambda \bar{\Lambda}$~\cite{Lambdapair} with an analysis that used
a $40.5~\mathrm{pb}^{-1}$ data sample that was collected 
at four different energy scan points during 2011 and 2012. 
The lowest energy point is 2.2324 GeV, only 1 MeV
above the $\Lambda\bar{\Lambda}$-threshold. 
These data made it possible to measure the Born cross section 
very near threshold.  To use the data as efficiently as possible, 
both events where $\Lambda$ and $\bar{\Lambda}$ decayed to 
the charged mode ($\mathrm{Br}(\Lambda \rightarrow p\pi^-) = 64\%$)
and events where the $\bar{\Lambda}$ decayed to the neutral mode
($\mathrm{Br}(\bar{\Lambda} \rightarrow \bar{n} \pi^0) = 36\%$) were
selected. In the first case, the identification relied on finding two
mono-energetic charged pions with evidence for a $\bar{p}$-annihilation
in the material of the beam pipe or the inner wall of the tracking chamber. 
In the second case, the $\bar{n}$-annihilation was identified with a
multi-variate analysis of variables provided by the electromagnetic
calorimeter.  Additonally, a mono-energetic $\pi^0$ was reconstructed 
to fully identify this decay channel. 
For the higher energy points, only the charged decay modes of 
$\Lambda$ and $\bar{\Lambda}$ were reconstructed by
identifying all the charged tracks and using the event kinematics.
The resulting measurement~\cite{Lambdapair} of the Born cross section 
are shown in Fig.~\ref{lambda}(a) together with previous 
measurements~\cite{DM2BB, LambdaBaBar}.
The Born cross section near threshold is found to be 
$312 \pm 51(\rm stat.) ^{+72}_{-45}$(\rm sys.) pb.
This result confirms BaBar's measurement~\cite{LambdaBaBar} 
but with much higher momentum transfer squared accuracy. 
Since the Coulomb factor is equal to 1 for neutral baryon pairs, 
the cross section is expected to go to zero at threshold. 
Therefore the observed threshold enhancement implies the existence 
of a complicated underlying physics scenario. 
The unexpected features of baryon pair production near threshold have driven 
a lot of theoretical studies, including scenarios that invoke bound states 
or unobserved meson resonances~\cite{Dalkarov, TheoRes, Xiao:2019qhl}. 
It was also interpreted as an attractive Coulomb interaction 
on the constituent quark level~\cite{Resumfct1, Resumfct2}. 
Another possible explanation is the final-state interactions which 
play an important role near the threshold~\cite{FSinter1, FSinter2, FSinter3}.
The BESIII measurement improves previous results at low invariant masses 
at least by 10$\%$ and even more above 2.4 GeV/c. 
The $\Lambda$ effective form factor extracted from 
the cross section measurement is shown in Fig.~\ref{lambda}(b).

\begin{figure}[!htbp]
\begin{center}
\includegraphics[width=0.49\textwidth]{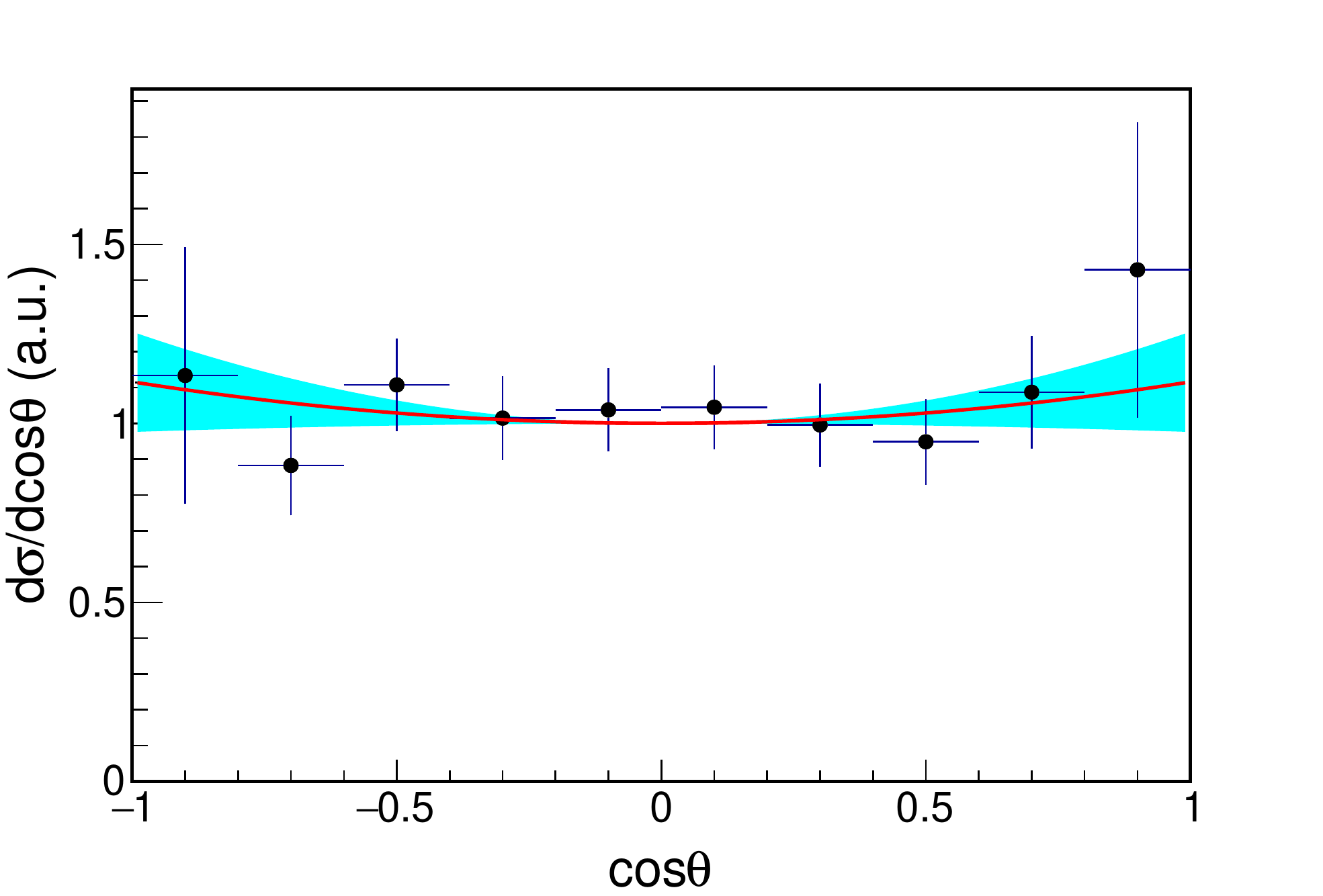}
\put(-190, 120){(a)}
\includegraphics[width=0.49\textwidth]{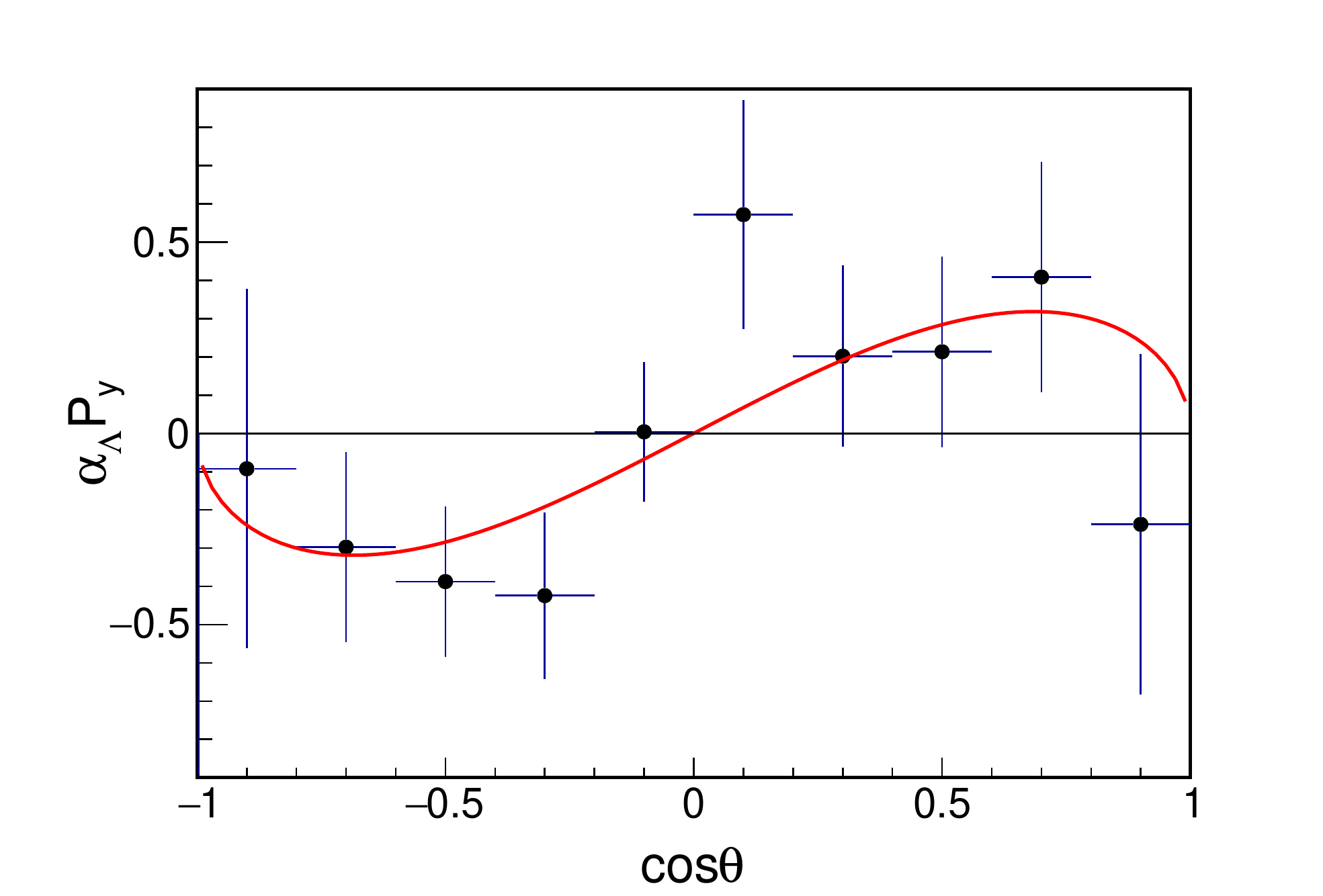}
\put(-180, 120){(b)}
\caption{(a) The acceptance corrected $\Lambda$ scattering angle 
distribution for $e^+e^{-} \rightarrow \Lambda \bar{\Lambda}$ at
$M_{\Lambda \bar{\Lambda}} = 2.396$ GeV.
(b) The product of the $\Lambda$ decay parameter $\alpha_\Lambda$ 
and $\Lambda$ polarization $P_y$ as a function of the scattering angle.
Plots are from Ref.~\cite{LambdaFF}.}
\label{cosThetaLambda}
\end{center}
\end{figure}

According to the optical theorem, there is a nonzero relative phase
between $G_E$ and $G_M$.
At $M_{\Lambda \bar{\Lambda}} = 2.396$ GeV, where we have 
the largest $\Lambda \bar{\Lambda}$ sample of 555 events 
from 66.9 pb$^{-1}$ data, a multidimensional analysis 
was used to make a full determination of the $\Lambda$ 
electromagnetic form factors for the first time for any baryon; 
the relative phase difference is 
$\Delta\Phi = 37^\circ \pm 12^\circ \pm 6^\circ$~\cite{LambdaFF} 
with the input parameter $\alpha_\Lambda = 0.750\pm0.010$ 
measured from $J/\psi$ decays~\cite{LambdaPolar}.
The improved determination of $\alpha_\Lambda$ also has profound 
implications for the baryon spectrum, since fits to such observables 
by theoretical models are a crucial element in determining 
the light baryon resonance spectrum, which provides a point of 
comparison for theoretical approaches~\cite{alphaLambda}.
The $|G_E/G_M|$ ratio was determined to be 
$R = 0.96 \pm 0.14(\rm stat.) \pm 0.02(\rm sys.)$ and 
the effective form factor at $M_{\Lambda \bar{\Lambda}} = 2.396$ GeV
was determined to be
$|G_{eff}| = 0.123 \pm 0.003 (\rm stat.) \pm 0.003 (\rm sys.)$.
The $\Lambda$ angular distribution and the polarization 
as a function of the scattering angle are shown in 
Fig.~\ref{cosThetaLambda}(a) and (b), respectively.
This first complete measurement of the hyperon electromagnetic 
form factor is a milestone in the study of hyperon structure, 
while the long-term goal is to describe charge and magnetization 
densities of the hyperons.

\subsection{The $\Lambda_c$ charmed baryon}
\begin{figure}[t]
\begin{center}
\begin{minipage}{0.45\linewidth}
\centerline{\includegraphics[width=1\linewidth,height=0.75\linewidth]{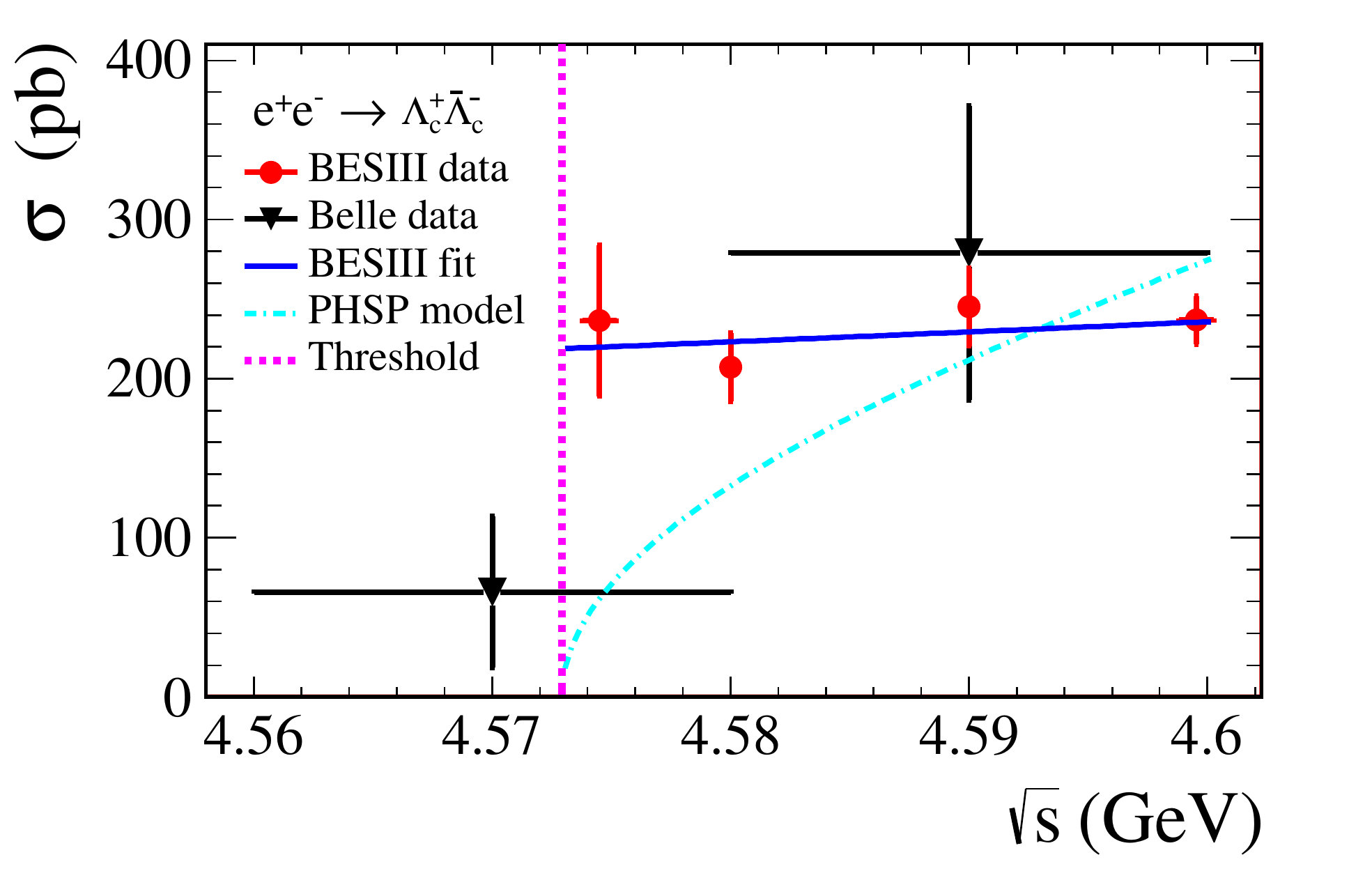}}
\end{minipage}
\put(-45,55){(a)}
\begin{minipage}{0.46\linewidth}
\centerline{\includegraphics[width=1\linewidth,height=0.75\linewidth]{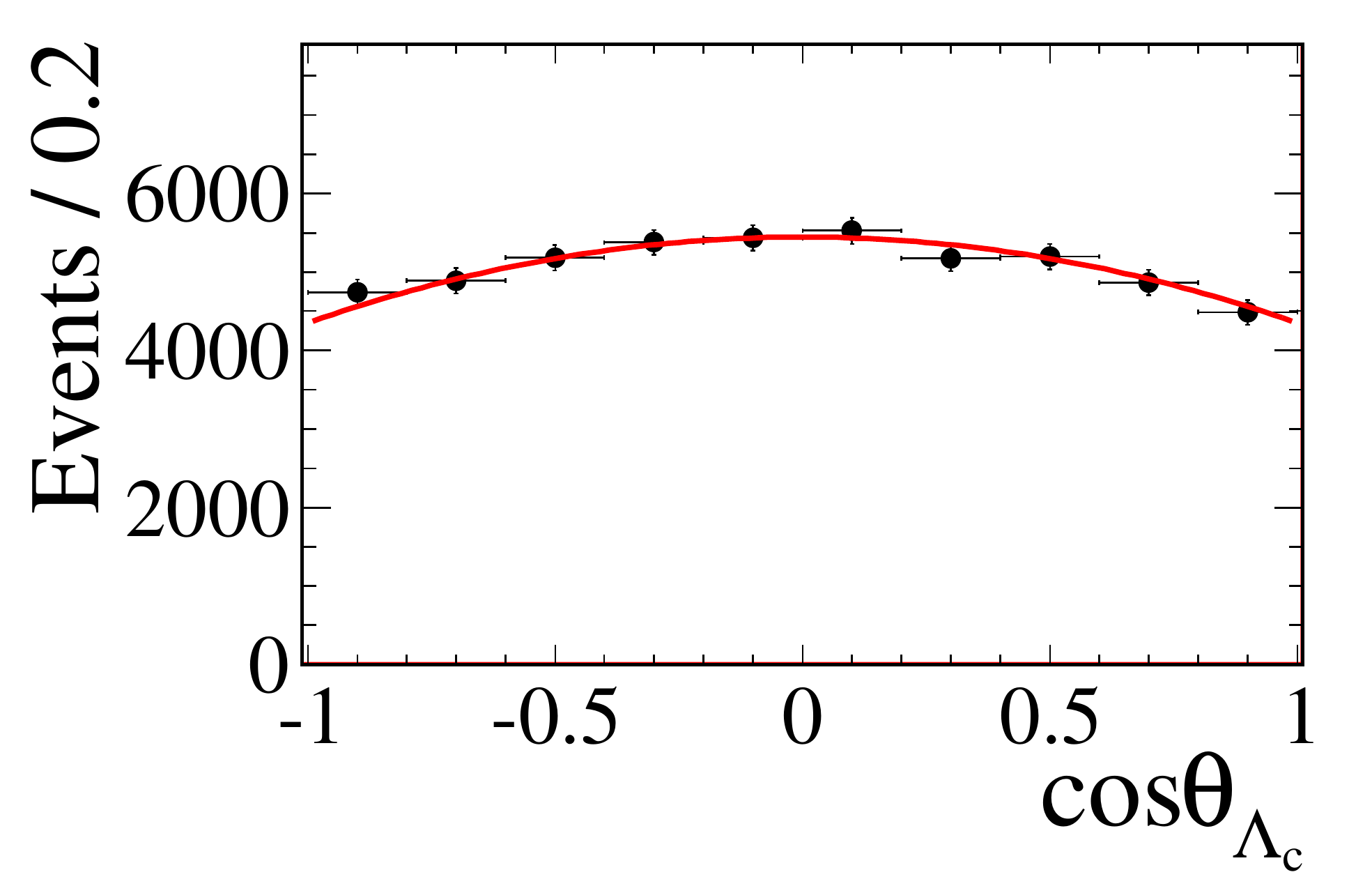}}
\end{minipage}
\put(-40,55){(b)}
\end{center}
\caption[]{The Born cross section of $e^+e^- \to \Lambda_c^+ \bar{\Lambda}_c^-$
obtained by BESIII and Belle (a).
The angular distribution and corresponding fit results in data at
$\sqrt{s} = 4.5995$ GeV (b). Plots are from Ref.~\cite{Lambdacpair}.}
\label{lambdac2}
\end{figure}

Experimental studies on the charmed baryons have been rather sparse.
The only previous study of process $e^+e^- \to \Lambda_c^+ \bar{\Lambda}_c^-$
is from the Belle experiment, which measured the cross section 
using ISR technique~\cite{LambdacBelle}, and reported a lineshape 
that implied the existence of a likely resonance, called the Y(4660).
Based on $631.3~\mathrm{pb}^{-1}$ data collected in 2014 at four 
energy points $\sqrt{s} = $4.5745, 4.5809, 4.5900 and 4.5995 GeV,
BESIII measured the $\Lambda_c^+ \bar{\Lambda}_c^-$ cross section
with unprecedented precision~\cite{Lambdacpair}.
The lowest energy point is only 1.6 MeV above the
$\Lambda_c^+ \bar{\Lambda}_c^-$ threshold.
At each of the energy points, ten Cabibbo-favored hadronic decay modes,
$\Lambda_c^+ \rightarrow p K^-\pi^+$, $p K_S^0$, $\Lambda \pi^+$,
$p K^-\pi^+ \pi^0$, $p K^0 \pi^0$, $\Lambda \pi^+ \pi^0$,
$p K_S \pi^+ \pi^-$, $\Lambda \pi^+ \pi^+ \pi^-$, $\Sigma^0 \pi^+$,
and $\Sigma^+\pi^+\pi^-$, as well as the corresponding charge-conjugate
modes were studied.
The total Born cross section is obtained from the weighted average
of the 20 individual measurements, and
the results are shown in Fig.~\ref{lambdac2}(a). 
Similar to the case for $e^+e^- \to p \bar{p}$, an abrupt rise 
in the cross-section just above threshold that is much steeper 
than phase-space expectations is discerned, which was not seen by
Belle due to limitations of the ISR method.
BESIII's measured cross section lineshape is different from Belle's, 
disfavoring a resonance like Y(4660) in the $\Lambda_c^+ \bar{\Lambda}_c^-$ 
channel. The BESIII results have driven discussions in the 
theoretical literature~\cite{Dai:2017fwx}.

High statistic data samples at $\sqrt{s} = $ 4.5745 and 4.5995 GeV
enabled studies of the polar angular distribution of $\Lambda_c$ in the
$e^+e^-$ center-of-mass system. The shape function $f(\theta) \propto
(1 + \alpha_{\Lambda_c}\cos^2\theta)$ is fitted to the combined data contaning
the yields of $\Lambda_c^+$ and $\bar{\Lambda}_c^-$ for all ten decay modes
as shown in Fig.~\ref{lambdac2}(b). 
The ratio between the electric and magnetic form factors $|G_E/G_M|$ 
can be extracted using
$|G_E/G_M|^2(1-\beta^2) = (1 - \alpha_{\Lambda_c})/ (1 + \alpha_{\Lambda_c})$.
From these distributions, the ratios $|G_E/G_M|$ of $\Lambda_c^+$ 
have been extracted for the first time: they are
$1.14 \pm 0.14~(\rm stat.) \pm 0.07~(\rm sys.)$ and 
$1.23 \pm 0.05~(\rm stat.) \pm 0.03~(\rm sys.)$ 
at $\sqrt{s} = $ 4.5745 and 4.5995 GeV, respectively.    



\section{Baryon challenges at BESIII}
\begin{figure}[htbp]
\begin{center}
\includegraphics[width=0.8\textwidth]{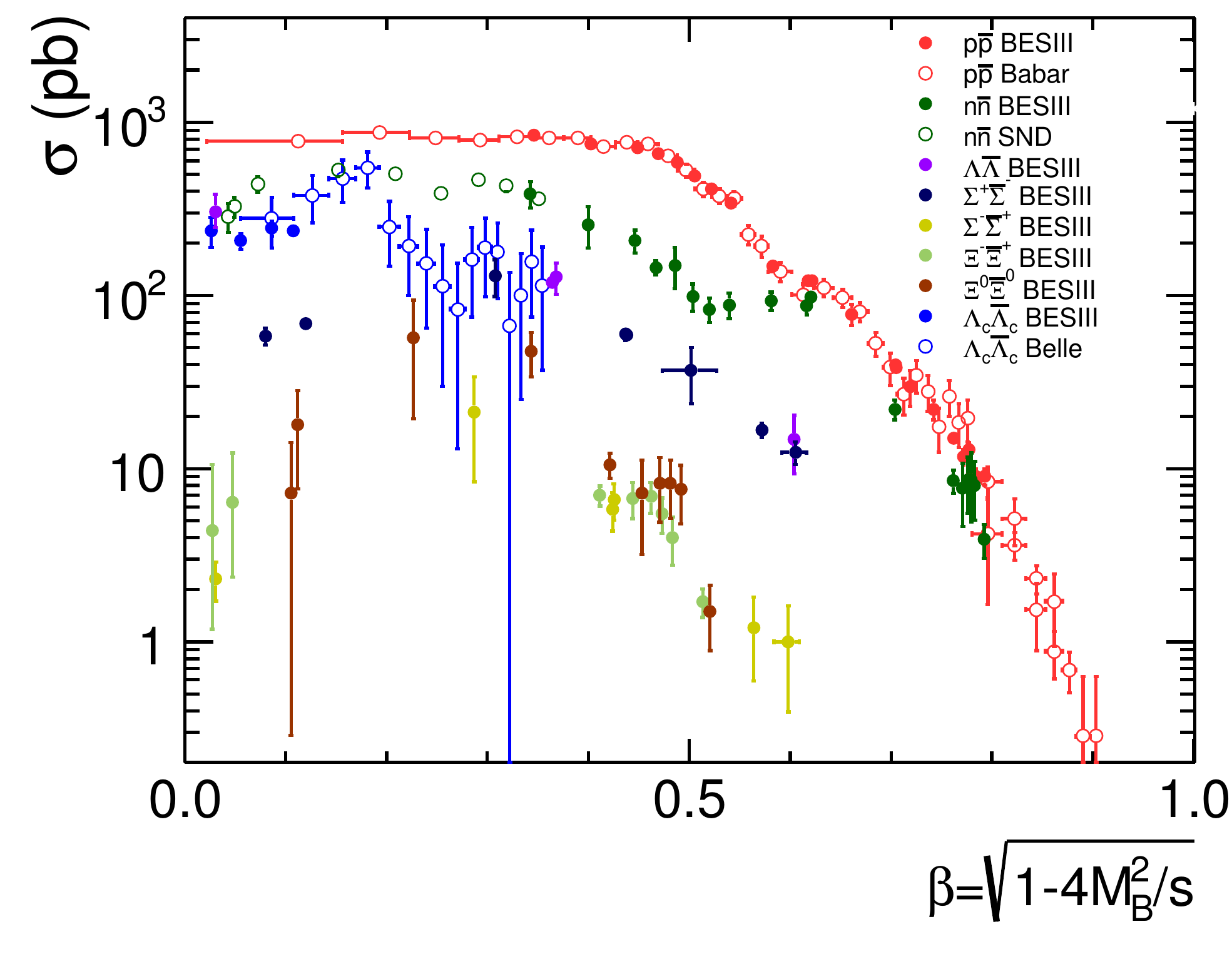}
\caption{A compilation of cross sections, revealing similar pattern 
for all $B \bar{B}$ pairs measured so far: 
$p\bar{p}$ by BaBar~\cite{ppbarBaBar1,ppbarBaBar2} and BESIII~\cite{ppbar2020}, 
$n\bar{n}$ by SND~\cite{SND2014, SNDnn} and BESIII~\cite{nnbar},
$\Lambda\bar{\Lambda}$ by BESIII~\cite{Lambdapair},
$\Sigma^{+}\bar{\Sigma}^{-}$/$\Sigma^{-}\bar{\Sigma}^{+}$ 
by BESIII~\cite{chargedSigma},
$\Xi^{-}\bar{\Xi}^{+}$/$\Xi^{0}\bar{\Xi}^{0}$ 
by BESIII~\cite{chargedXi, neutralXi},
$\Lambda_c^+\bar{\Lambda}_c^-$ by Belle~\cite{LambdacBelle} 
and BESIII~\cite{Lambdacpair}.}
\label{xs_vs_beta}
\end{center}
\end{figure}

The energy thresholds for pair production of all of the ground state 
spin 1/2 SU(3) octet and spin 3/2 decuplet are accessible to BESIII. 
Baryon form factor measurements are among the most important reasons
why BESIII has collected an unprecedented amount of off-resonance data.
From the analysis of existing data, it is expected that the ratio of 
the absolute values of the $\Lambda$ electromagnetic form factors, 
$|G_E/G_M|$, can be measured at five energy points. 
The most interesting findings are the abrupt cross section jumps
at threshold followed by a nearly flat that has been 
observed for $\Lambda\bar{\Lambda}$, $\Lambda_c^+\bar{\Lambda}_c^-$, 
$p\bar{p}$, $n\bar{n}$, etc. 
If the BEPCII energy could be lowered to the vicinity of nucleon-antinucleon 
threshold, BESIII will be able to confirm the $p\bar{p}$ and $n\bar{n}$ 
cases with much better precision. 
Figure~\ref{xs_vs_beta} shows the cross section lineshapes
for a variety of baryon-antibaryon pairs, including those 
that were recently measured for singly-stranged
$\Sigma^{+}\bar{\Sigma}^{-}$/$\Sigma^{-}\bar{\Sigma}^{+}$~\cite{chargedSigma},
doubly-stranged $\Xi^{-}\bar{\Xi}^{+}$~\cite{chargedXi} and
$\Xi^{0}\bar{\Xi}^{0}$~\cite{neutralXi}.
They all seem to share the common feature with a plateau 
starting from the baryon-pair production threshold,
though for some channels ideally more statistics are needed.
The behavior of $\Sigma^{0}\bar{\Sigma}^{0}$ 
(the last member to be covered for the spin 1/2 SU(3) octet baryons) 
and other baryon-pairs will be reported in the near future.

\section{Summary and Prospects}
The measurements of baryon form factors have been an important 
ongoing activity at BESIII. Form factors of proton with the best
precision were obtained in the time-like region, and the electric 
form factor of proton was measured for the first time.
Measurements of the neurtron time-like form factor with 
unprecedented precision have also been reported.
The $\Lambda$ and $\Lambda_c$ were studied and in both cases 
abnormal cross section enhancements were observed 
near the production thresholds.
The form factors of the $\Lambda_c$ were extracted for the first time.

In addition $\Sigma^{+}/\Sigma^{-}$~\cite{chargedSigma}, 
$\Xi^{-}$~\cite{chargedXi} and $\Xi^{0}$~\cite{neutralXi} 
form factor measurements were recently reported, 
and results for $\Sigma^{0}$ will soon be released.
BESIII also has a plan to explore the nucleon production threshold
by taking data in $1.8-2.0$ GeV for $\sim 100$ pb$^{-1}$ at 23
enengy points~\cite{WhitePaper}, in order to study the anomalous threshold
cross section behavior in more detail.
With numerous first measurements and interesting discoveries, 
these studies shed new light for understanding 
the interactions and fundamental structure of particles.

It will take long time to ultimately unravel the fundamental structure 
of baryons. Further improvements in form factor measurement of baryons 
will continue to be the focus of future powerful electron-ion colliders 
in America (EiC)~\cite{EiC} and China (EicC)~\cite{EicC}, super
electron-positron colliders in China~\cite{STCF} and Russia~\cite{SCTF} 
for the space-like and time-like regions, respectively.


\begin{acknowledgments}
The authors thank Prof. S. L. Olsen for helpful suggestions and proofreading.
This work is supported in part by National Natural Science Foundation 
of China (NSFC) under Contract Nos. 12035013, 12061131003, 
11911530140 and 11335008;
Joint Large-Scale Scientific Facility Funds of the NSFC and 
the Chinese Academy of Sciences under Contract No. U1832103; 
National Key Research and Development Program of China 
under Contract No. 2020YFA0406403;
National Key Basic Research Program of China under Contract No. 2015CB856705.
Conflict of interest statement: none declared.
\end{acknowledgments}


\end{document}